\documentclass[twocolumn,prb]{revtex4-2}%
\usepackage{amsmath}
\usepackage{amssymb}
\usepackage{graphicx}
\usepackage{txfonts}
\usepackage[titletoc]{appendix}
\usepackage{array}
\usepackage{mathrsfs}
\usepackage{subfigure,overpic}
\usepackage{dcolumn}
\usepackage{epstopdf}
\usepackage{braket}
\usepackage[colorlinks,linkcolor=blue, urlcolor=blue,
anchorcolor=blue,citecolor=blue,breaklinks=true]{hyperref}
\usepackage{amsfonts}%
\providecommand{\U}[1]{\protect\rule{.1in}{.1in}}
\providecommand{\U}[1]{\protect\rule{.1in}{.1in}}
\begin{document}
\title{Anisotropy-induced collapse of Landau levels in Weyl semimetals and its detection via the planar Hall effect}
\author{Fu-Yang Chen}
\author{Zhuo-Hua Chen}
\author{Hou-Jian Duan}
\author{Mou Yang}
\author{Rui-Qiang Wang}
\email{wangruiqiang@m.scnu.edu.cn}
\author{Ming-Xun Deng}
\email{dengmingxun@scnu.edu.cn}
\affiliation{$^{1}$Guangdong Basic Research Center of Excellence for Structure and
Fundamental Interactions of Matter, Guangdong Provincial Key Laboratory of
Quantum Engineering and Quantum Materials, School of Physics, South China
Normal University, Guangzhou 510006, China}
\affiliation{$^{2}$Guangdong-Hong Kong Joint Laboratory of Quantum Matter, Frontier
Research Institute for Physics, South China Normal University, Guangzhou
510006, China}

\begin{abstract}
The planar Hall effect (PHE) is a powerful tool for characterizing Weyl
semimetals (WSMs). Here, we investigate the PHE in general anisotropic WSMs
under strong magnetic fields. We analytically derive the Landau levels (LLs)
and their wavefunctions using the Bogoliubov transformation, where the tilt
vector, anisotropic axis of the Fermi velocity, and the magnetic field can be
oriented in arbitrary directions. Notably, due to the interaction with the
magnetic field and the anisotropy of the Fermi velocity, the component of the
tilt vector perpendicular to the magnetic field can induce a tilt in the LLs
parallel to the magnetic field. Our analytical results show that the LLs do
not collapse in type-\textrm{I} WSMs but must collapse in type-\textrm{II}
WSMs when the magnetic field is vertical to the tilt vector. More importantly,
we demonstrate that the magnetotransport signal of the LL collapse, which
manifests as significant enhancement and quantum oscillations in the
longitudinal and planar Hall conductivities simultaneously, can be used to
identify the phase transition from type-\textrm{I} to type-\textrm{II} WSMs.

\end{abstract}
\maketitle

\section{INTRODUCTION}

Massless Weyl fermions, originally predicted in high-energy physics as
solutions to the Weyl equation\cite{Weyl56}, emerge as low-energy
quasiparticles in a class of topological materials known as Weyl semimetals
(WSMs)\cite{Xu349,Lv5}. WSMs provide a novel platform for exploring
relativistic quantum phenomena in condensed matter
systems\cite{Wan83,Burkov107,Zhang92,Wang123,Li125}. In WSMs, the conduction
and valence bands intersect linearly at discrete points referred to as Weyl
nodes, which carry quantized topological charge (or chirality). This
topological protection makes WSMs robust against symmetry-preserving
perturbations and ideal for investigating topological phases of
matter\cite{Lv5}. The intrinsic chirality of Weyl nodes underpins many of the
unique phenomena observed in WSMs, including the chiral anomaly, Fermi arc
surface states, and unconventional magnetotransport behavior in external
electromagnetic
fields\cite{Wan83,Ojanen87,Huang5,Zhang7,Jia15,Burkov27,Armitage90}.

When subjected to a strong magnetic field, the unique topology of WSMs gives
rise to quantized Landau levels (LLs), leading to phenomena such as
Shubnikov-de Haas (SdH) oscillations\cite{L126}. The SdH oscillations,
typically observed through photon absorption or magnetic field-dependent
transport measurements, serve as powerful tools for probing the Fermi surface
geometry, effective masses of charge carriers, and Berry phase effects in
these materials\cite{Armitage90,Orlita10,Xiong350,Liang14}. The LL structure
in WSMs differs significantly from that in conventional nonrelativistic
electron systems. Notably, the chiral zeroth LLs in WSMs will induce the
chiral anomaly, which results in positive longitudinal
magnetoconductivity\cite{Liang3,Cichorek13,Yang11,Shekhar11,Li6} and unique
quantum oscillations\cite{Deng122}. Another intriguing magnetotransport
phenomenon in WSMs is the planar Hall effect (PHE), where a transverse current
appears coplanar with the applied magnetic and electric
fields\cite{Taskin8,Burkov96,Nandy119,Deng99}. The giant PHE in isotropic WSMs
is closely related to the chiral anomaly\cite{Burkov96}.

Recent studies have demonstrated that the magnetoconductivity in anisotropic
WSMs exhibits a strong dependence on the polar or azimuthal angle of the
anisotropy axis\cite{Jiang126,Sato132,Dotdaev108,Balduini109,Li97}. The giant
PHE has been experimentally observed in both type-\textrm{I} WSMs (such as
\textrm{GdPtBi} and \textrm{Cd}$_{3}$\textrm{As}$_{2}$%
)\cite{Li97,Nitesh98,Wu98,Li98} and type-\textrm{II} WSMs (including
\textrm{MoTe}$_{2}$ and \textrm{VAl}$_{3}$%
)\cite{Chen98,Singha98,Li100,Meng32,Vashist}. On the other hand, LLs in WSMs
can collapse when subjected to external electric fields\cite{Arjona96} or
strain\cite{Lee106}. More intriguingly, the collapse of the LLs is strongly
dependent on the direction of the applied external fields. For example, in
type-\textrm{II} WSMs, the LLs remain intact when electric and magnetic fields
are parallel, while they invariably collapse when these fields are
perpendicular, regardless of the magnitude of the external
fields\cite{Yu117,Serguei117,Udagawa117}. These observations have motivated
various theoretical models to explain the underlying
mechanisms\cite{Burkov96,Nandy119,Deng99,Ma99,Wei107,Onofre108,Wang5}. In
tilted WSMs, the Lorentz boost approach has been developed to solve analytical
expression for the LLs\cite{Serguei117}, similar to the case of an isotropic
WSM subjected to crossed electric and magnetic fields\cite{Deng99}. However,
in the presence of multiple anisotropy axes, finding an appropriate Lorentz
boost becomes challenging. Moreover, the Lorentz boost approach does not yield
an explicit expression for the normalized wavefunction. Additionally, while
the frequency spectrum of photoconductivity has been proposed as a means to
observe LL collapse\cite{Yu117}, detecting this phenomenon through
magnetotransport remains rarely reported. Although it is known that LL
collapse depends on the geometry of the Fermi surface, its relationship to the
phase transition from type-\textrm{I} to type-\textrm{II} WSMs has not been fully
explored.

In this work, by using the approach of the Bogoliubov transformation, we
analytically derive both the LLs and wavefunctions for general anisotropic
WSMs, where the tilt vector and anisotropic axis of the Fermi velocity, as
well as the magnetic field, can be oriented in arbitrary directions. Our
analytical expressions show that the LL collapse is closely related to the
phase transition of WSMs. Specifically, the LLs remain intact in
type-\textrm{I} WSMs, but collapse inevitably in type-\textrm{II} WSMs for
certain magnetic field orientations. Since the LLs in anisotropic WSMs are
highly sensitive to the direction of the magnetic field, we propose detecting
the LL collapse via the PHE. We find that the longitudinal conductivity and
planar Hall conductivity exhibit significant enhancement simultaneously at the
boundary of LL collapse, which are decorated with quantum oscillations and can
be used to experimentally detect the phase transition from type-\textrm{I} to
type-\textrm{II} WSMs. The rest of this paper is organized as follows. In Sec.
\ref{CLW}, we first derive the LLs and wavefunctions analytically, and then
demonstrate the relation between the LL collapse and the phase transition from
type-\textrm{I} to type-\textrm{II} WSMs. The PHE associated with the LL
collapse is discussed in Sec. \ref{PFC}. The last section contains a concise summary.

\begin{figure}[ptb]
\centering \includegraphics[width=0.46\textwidth]{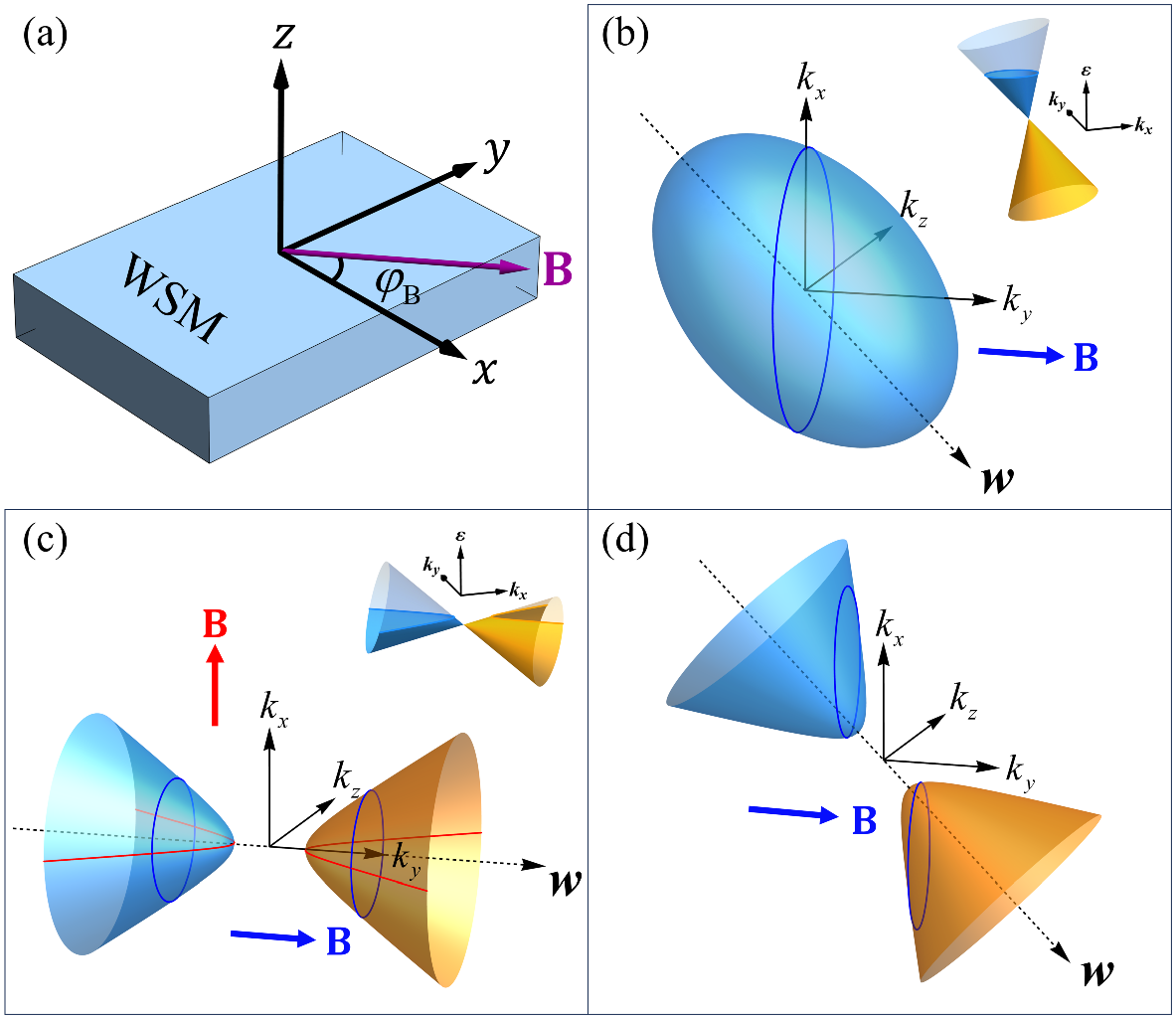}\caption{(a)
Schematic illustration of a WSM subjected to strong magnetic fields. (b) The
elliptical Fermi surface for type-\textrm{I} WSMs and (c)-(d) the
paraboloid-shaped Fermi surface for type-\textrm{II} WSMs. The inset in the
upper right corner of (b)-(c) demonstrates the Weyl cone and Fermi pockets at
$k_{z}=0$. The blue and red curves on the Fermi surfaces indicate their
cross-section perpendicular to the magnetic field $\boldsymbol{B}$ represented
by the red and blue arrows. In type-\textrm{II} WSMs, when $\boldsymbol{B}%
\cdot\boldsymbol{w}=0$, the cross-section of the Fermi surface is open,
demonstrated by the red curves in (c), while for $\boldsymbol{B}%
\cdot\boldsymbol{w}\neq0$, the cross-section can remain closed, illustrated by
the blue curves in (d).}%
\label{Fig1}%
\end{figure}

\section{Collapse of LLs in anisotropic WSMs}

\label{CLW}

We start from the low-energy effective Hamiltonian for anisotropic
WSMs\cite{Lee106,Dotdaev108,Serguei117}%
\begin{equation}
\mathcal{H}\left(  \boldsymbol{k}\right)  =\hbar\hat{\boldsymbol{\upsilon}%
}\cdot\boldsymbol{k}, \label{eq_H1}%
\end{equation}
where $\boldsymbol{k}$ denotes the wavevector, $\hat{\upsilon}_{i=x,y,z}%
=w_{i}+\upsilon_{i}\sigma_{i}$ represents the velocity operator, and
$\sigma_{i}$ is the Pauli matrix. The WSMs described by Eq. (\ref{eq_H1})
contain two anisotropy principal axes along $\boldsymbol{w}=(w_{x},w_{y}%
,w_{z})$ and $\boldsymbol{v}=(\upsilon_{x},\upsilon_{y},\upsilon_{z})$,
respectively, which are generally not collinear. The Weyl fermion excitations
possess definite chirality defined as $\chi=\mathrm{sgn}(\upsilon_{x}%
\upsilon_{y}\upsilon_{z})$, and the continuum excitation spectrum is given by%
\begin{equation}
\varepsilon_{\boldsymbol{k}\pm}=\hbar\boldsymbol{w}\cdot\boldsymbol{k}\pm
\hbar\sqrt{\upsilon_{x}^{2}k_{x}^{2}+\upsilon_{y}^{2}k_{y}^{2}+\upsilon
_{z}^{2}k_{z}^{2}}. \label{eq_H2}%
\end{equation}
As indicated in Eq. (\ref{eq_H2}), a finite $\boldsymbol{w}$ will tilt the
Weyl cone. The tilt vector $\boldsymbol{w}$ can not open gap for the spectrum
and thus does not alter the topology of the energy band. However, since the
topology of a WSM is characterized by the Chern number enclosed by the Fermi
surface, the tilt vector can induce the phase transition from type-\textrm{I}
to type-\textrm{II }WSMs through modifying the geometry of the Fermi surface.
Here, the boundary of the phase transition is determined by $t_{w}=1$ where
\begin{equation}
t_{w}=\sqrt{\frac{w_{x}^{2}}{\upsilon_{x}^{2}}+\frac{w_{y}^{2}}{\upsilon
_{y}^{2}}+\frac{w_{z}^{2}}{\upsilon_{z}^{2}}}.
\end{equation}
In type-\textrm{I} WSMs, $t_{w}<1$ and the Fermi surface\ is a closed
ellipsoid, see Fig. \ref{Fig1} (b). Conversely, in type-\textrm{II} WSMs where
$t_{w}>1$, the Weyl cone becomes over-tilted, and the Fermi surface turns to
two open hyperboloids, including both particle and hole Fermi pockets, as
shown in Fig. \ref{Fig1} (c).

When the WSM is subjected to a magnetic field, the Hamiltonian can be obtained
from Eq. (\ref{eq_H1}) through the Peierls substitution $\boldsymbol{k}%
\rightarrow\boldsymbol{k}+e\boldsymbol{A}/\hbar$. Without loss of generality
and for the convenience of discussion, we fix the magnetic field in the
$x$-$y$ plane, represented by the vector potential $\boldsymbol{A}=B\left(
0,0,y^{\prime}\right)  $, where $y^{\prime}=y\cos\varphi_{B}-x\sin\varphi_{B}$
and $\boldsymbol{B}\equiv\boldsymbol{\nabla}\times\boldsymbol{A}=B\left(
\cos\varphi_{B},\sin\varphi_{B},0\right)  $, as shown in Fig. \ref{Fig1} (a).
When the magnetic field is included, $[k_{x(y)},\mathcal{H}\left(
\boldsymbol{k}+e\boldsymbol{A}/\hbar\right)  ]\neq0$ and therefore, $k_{x(y)}$
itself is no longer conserved. However, $k_{x}^{\prime}=k_{x}\cos\varphi
_{B}+k_{y}\sin\varphi_{B}$ remains a good quantum number because $[y^{\prime
},k_{x}^{\prime}]=0$. Consequently, we can perform a unitary transformation
and express the Hamiltonian in terms of $k_{x}^{\prime}$ and $y^{\prime}$ as
$\mathcal{\tilde{H}}_{\boldsymbol{k}}=U^{-1}\mathcal{H}\left(  \boldsymbol{k}%
+e\boldsymbol{A}\mathbf{/}\hbar\right)  U$, which can be derived as
\begin{equation}
\mathcal{\tilde{H}}_{\boldsymbol{k}}=\hbar(w_{x}^{\prime}k_{x}^{\prime}%
+\omega_{-}\hat{\Pi}+\omega_{+}\hat{\Pi}^{\dag})+\hbar\left(
\begin{array}
[c]{cc}%
\gamma\upsilon_{B}k_{x}^{\prime} & \sqrt{2}\omega_{c}\hat{\Pi}\\
\sqrt{2}\omega_{c}\hat{\Pi}^{\dag} & -\gamma\upsilon_{B}k_{x}^{\prime}%
\end{array}
\right)  . \label{eq_Hp}%
\end{equation}
Here, the operators $\hat{\Pi}=\frac{s_{z}\xi+\partial_{\xi}}{\sqrt{2}%
}-i\alpha\frac{\upsilon_{B}k_{x}^{\prime}}{\sqrt{2}\omega_{c}}$ and $\hat{\Pi
}^{\dag}=\frac{s_{z}\xi-\partial_{\xi}}{\sqrt{2}}+i\alpha\frac{\upsilon
_{B}k_{x}^{\prime}}{\sqrt{2}\omega_{c}}$ satisfy the commutation relation
$[\hat{\Pi},\hat{\Pi}^{\dag}]=s_{z}$, in which $s_{z}=\mathrm{sgn}%
(\upsilon_{z})$, $\xi=\sqrt{\frac{|\upsilon_{z}|}{\upsilon_{B}}}%
\frac{y^{\prime}+\ell_{B}^{2}k_{z}}{\ell_{B}}$, $\omega_{c}=\sqrt
{|\upsilon_{z}|\upsilon_{B}}/\ell_{B}$, and $\ell_{B}=\sqrt{\hbar/|eB|}$
denotes the magnetic length. The unitary matrix, given by%
\begin{equation}
U=\frac{1}{\sqrt{2}}\left(
\begin{array}
[c]{cc}%
e^{-i\tilde{\varphi}_{B}} & e^{-i\tilde{\varphi}_{B}}\\
1 & -1
\end{array}
\right)
\end{equation}
with $\tan\tilde{\varphi}_{B}=\frac{\upsilon_{x}}{\upsilon_{y}}\tan\varphi
_{B}$, is constructed from the wavefunctions of $\tilde{\sigma}_{x}=\sigma
_{x}\cos\tilde{\varphi}_{B}+\sigma_{y}\sin\tilde{\varphi}_{B}$. For brevity,
we noted $\gamma=\upsilon_{x}\upsilon_{y}/\upsilon_{B}^{2}$, $\alpha
=\frac{\upsilon_{x}^{2}-\upsilon_{y}^{2}}{2\upsilon_{B}^{2}}\sin2\varphi_{B}$,
$\upsilon_{B}=|\upsilon_{y}\cos\varphi_{B}/\cos\tilde{\varphi}_{B}|$,
$w_{x}^{\prime}=w_{x}\cos\varphi_{B}+w_{y}\sin\varphi_{B}+\alpha w_{y}%
^{\prime}$, $w_{y}^{\prime}=w_{y}\cos\varphi_{B}-w_{x}\sin\varphi_{B}$, and%
\begin{equation}
\omega_{\pm}=\frac{\omega_{c}}{\sqrt{2}}\left(  \frac{w_{z}}{|\upsilon_{z}%
|}\pm i\frac{w_{y}^{\prime}}{\upsilon_{B}}\right)  .
\end{equation}
\begin{figure}[ptb]
\centering \includegraphics[width=0.49\textwidth]{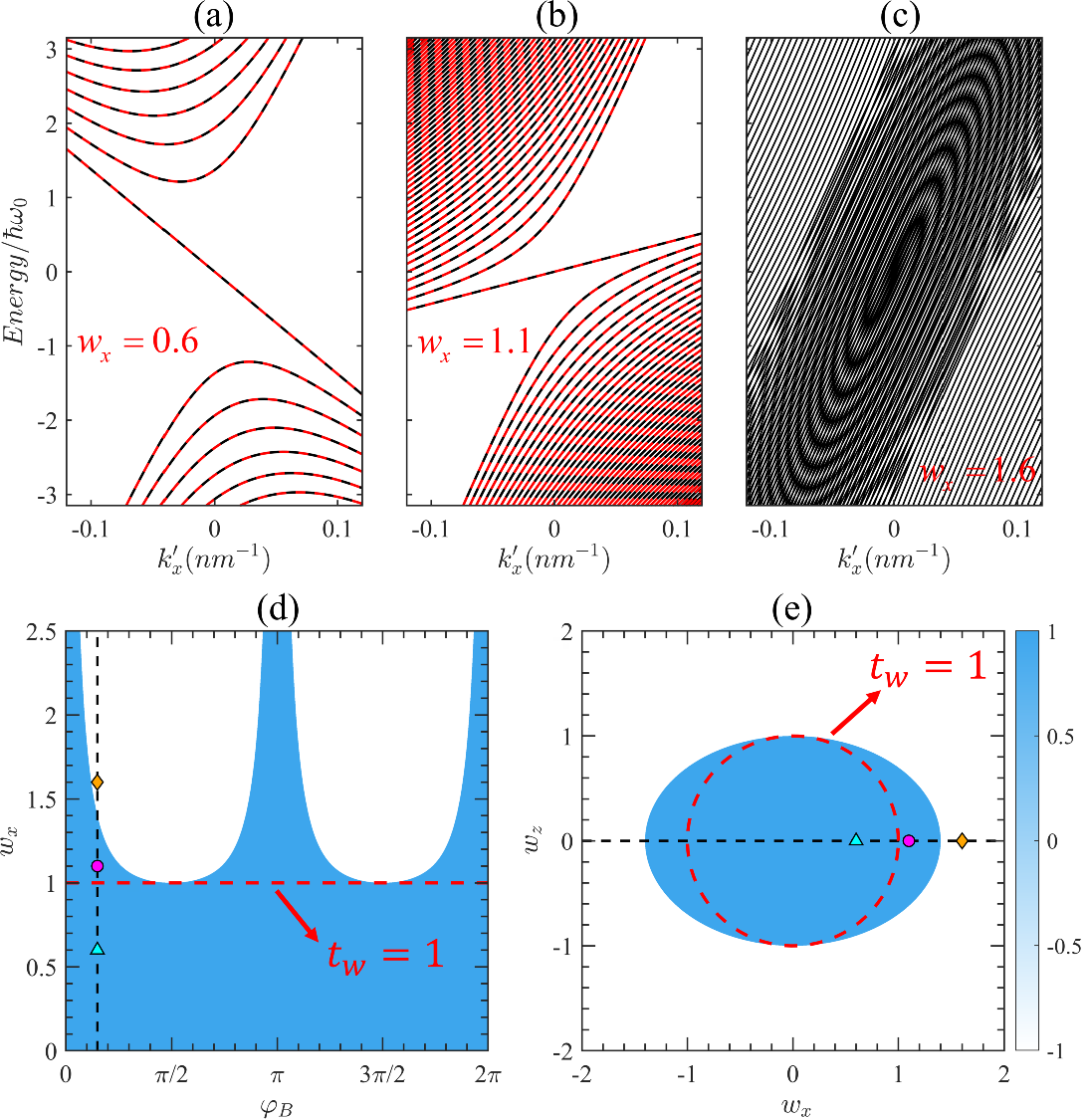}\caption{(a)-(c)
Evolution of the LLs with respect to $w_{x}$, where the dark and red-dashed
curves represent the numerical and analytical results, respectively. (d)-(e)
Distribution of $\mathrm{sgn}(1-2\omega_{+}\omega_{-}/\omega_{c}^{2})$ in
parameter space with (d) $w_{z}=0$ and (e) $\varphi_{B}=0.15\pi$, where the
LLs at the green-filled triangle, red-filled circle and brown-filled diamond
points are presented in (a)-(c), respectively. In the blue regions, the LLs
are well defined and in the white regions, the LLs collapse. The red dashed
curves represent the boundary of the phase transition between type-\textrm{I}
and type-\textrm{II} WSMs. Here, $\omega_{0}=\sqrt{|\upsilon_{y}\upsilon_{z}%
|}/\ell_{B}$ is chosen as the unit of energy, and the rest parameters are
$B=1$ $\mathrm{T}$, $w_{y}=0$, $\upsilon_{y}=0.5$, and $\upsilon_{x}%
=\upsilon_{z}=1$.}%
\label{Fig2}%
\end{figure}As distinct from the tilted or electrically-driven WSMs subjected
to strong magnetic fields reported in
Refs.\cite{Yu117,Goerbig78,Deng99,Peres19,Alisu107}, the coefficients in front
of $\hat{\Pi}$ and$\ \hat{\Pi}^{\dag}$, because of the tilt vector, are
different in the diagonal term of Eq. (\ref{eq_Hp}). At the same time, an
additional term $\propto\alpha k_{x}^{\prime}$ appears in the off-diagonal
elements of Eq. (\ref{eq_Hp}) due to the anisotropy of the Fermi velocity.
These additional factors couple the original Hilbert subspace, complicating
the solution of the eigenvalue problem.

Numerically, it is convenient to project Eq. (\ref{eq_Hp}) onto the particle
number representation by the ladder operators $\hat{a}^{\dag}=(\xi
-\partial_{\xi})/\sqrt{2}$ and $\hat{a}=(\xi+\partial_{\xi})/\sqrt{2}$, which
allows us to express the Hamiltonian as%
\begin{equation}
\mathcal{H}^{\prime}\left(  \boldsymbol{k}\right)  =\sum_{p=0}^{\infty}\left[
c_{p}^{\dag}h_{p}c_{p}+\left(  c_{p}^{\dag}h_{p,p+1}c_{p+1}+h.c.\right)
\right]  , \label{eq_HSQ}%
\end{equation}
where $c_{p}=(c_{p-1,s_{z}\uparrow},c_{p,s_{z}\downarrow})^{T}$, and
\begin{equation}
h_{p}=\hbar w_{x}^{\prime}k_{x}^{\prime}+s_{z}\left(
\begin{array}
[c]{cc}%
\gamma\hbar\upsilon_{B}k_{x}^{\prime} & \sqrt{2p}\hbar\omega_{c}\\
\sqrt{2p}\hbar\omega_{c} & -\gamma\hbar\upsilon_{B}k_{x}^{\prime}%
\end{array}
\right)  ,
\end{equation}%
\begin{equation}
h_{p,p+1}=\left(
\begin{array}
[c]{cc}%
\sqrt{p}\hbar\omega_{-} & 0\\
is_{z}\alpha\hbar\upsilon_{B}k_{x}^{\prime} & \sqrt{p+1}\hbar\omega_{-}%
\end{array}
\right)  .
\end{equation}
Here, $c_{p,s_{z}\downarrow}|0\rangle=0$, $c_{p^{\prime},s_{z}\downarrow
}|p\rangle=\Theta(p)|p-1\rangle\delta_{pp^{\prime}}$, $|p\rangle$ is the
eigenstate of $\hat{a}^{\dag}\hat{a}$, i.e., $\hat{a}^{\dag}\hat{a}%
|p\rangle=p|p\rangle$, and $\Theta(x)$ is the unit step function. By
diagonalizing Eq. (\ref{eq_HSQ}), we can obtain the discretized energy
spectrum. The evolution of the LLs with $w_{x}$ for $w_{y(z)}=0$ is
demonstrated in Figs. \ref{Fig2}(a)-(c). For $w_{x}<1$, the LLs are well
defined. With increasing $w_{x}$, the spacing between the LLs decreases
gradually. When $w_{x}>1$, the system transitions to a type-\textrm{II} WSM,
where the conduction and valence bands can overlap and couple to each other,
causing the LL collapse, as shown in Fig. \ref{Fig2}(c).

To clarify the collapse behavior of the LLs, we solve the eigenvalue problem
analytically. To this end, we perform the Bogoliubov transformation (see
Appendix \ref{DBT})%
\begin{equation}
\hat{\Pi}=\lambda_{+}\hat{\beta}+\lambda_{-}\frac{\omega_{+}}{\omega_{-}}%
\hat{\beta}^{\dag}-2\eta^{2}\lambda_{+}\lambda_{-}\frac{\omega_{\varepsilon}%
}{\omega_{-}}, \label{eq_Pi}%
\end{equation}
where $[\hat{\beta},\hat{\beta}^{\dag}]=1$, $\lambda_{\pm}=(\eta\pm s_{z}%
\eta^{-1})/2$, and
\begin{equation}
\eta=\left(  1-2\omega_{+}\omega_{-}/\omega_{c}^{2}\right)  ^{-1/4}=\left(
1-\frac{w_{y}^{\prime2}}{\upsilon_{B}^{2}}-\frac{w_{z}^{2}}{\upsilon_{z}^{2}%
}\right)  ^{-1/4}. \label{eq_RP}%
\end{equation}
Here, $\hbar\omega_{\varepsilon}=\varepsilon-\hbar w_{x}^{\prime}k_{x}%
^{\prime}$ and $\varepsilon$ is the energy corresponding to the wavefunction
$|\Psi\rangle=(|\psi\rangle,|\varphi\rangle)^{T}$. Then, by introducing the
auxiliary operator
\begin{equation}
\mathcal{\hat{F}}=\eta^{4}\omega_{\varepsilon}^{2}-\gamma^{2}\upsilon_{B}%
^{2}k_{x}^{\prime2}-\eta^{-2}\omega_{c}^{2}\left(  2\hat{\beta}^{\dag}%
\hat{\beta}+1\right)  , \label{eq_MF}%
\end{equation}
we find that the wavefunction satisfies the matrix equation
\begin{equation}
\left(
\begin{array}
[c]{cc}%
\mathcal{\hat{F}}-s_{z}\omega_{c}^{2} & -\sqrt{2}s_{z}\omega_{c}\omega_{+}\\
\sqrt{2}s_{z}\omega_{c}\omega_{-} & \mathcal{\hat{F}}+s_{z}\omega_{c}^{2}%
\end{array}
\right)  \left(
\begin{array}
[c]{c}%
|\psi\rangle\\
|\varphi\rangle
\end{array}
\right)  =0. \label{eq_AO}%
\end{equation}
According to Eq. (\ref{eq_AO}), the wavefunction can be written as (see
Appendix \ref{DFW})%
\begin{equation}
|\Psi_{n}\rangle=\sqrt{\lambda_{+}}\left(
\begin{array}
[c]{c}%
C_{n}|n-1\rangle\\
D_{n}|n\rangle
\end{array}
\right)  -\sqrt{\frac{\lambda_{-}}{\omega_{+}\omega_{-}}}\left(
\begin{array}
[c]{c}%
\omega_{+}D_{n}|n\rangle\\
\omega_{-}C_{n}|n-1\rangle
\end{array}
\right)  , \label{eq_WF}%
\end{equation}
in which $\hat{\beta}|n\rangle=\sqrt{n}|n-1\rangle$ and $\hat{\beta}^{\dag
}|n\rangle=\sqrt{n+1}|n+1\rangle$. Subsequently, by substituting Eq.
(\ref{eq_Pi}) and (\ref{eq_WF}) into the eigenequation of the Hamiltonian
(\ref{eq_Hp}), i.e., $\mathcal{\tilde{H}}_{\boldsymbol{k}}|\Psi_{n}%
\rangle=\varepsilon_{n}|\Psi_{n}\rangle$, we can determine $C_{n}=s_{n}%
R_{n,+}$ and $D_{n}=R_{n,-}$, with $s_{n}=\mathrm{sgn}(n)$ and
\begin{equation}
R_{n,\pm}=\left(  \eta^{-1}\pm\frac{s_{n}s_{z}\gamma\upsilon_{B}k_{x}^{\prime
}}{\sqrt{2|n|\omega_{c}^{2}+\eta^{2}\gamma^{2}\upsilon_{B}^{2}k_{x}^{\prime2}%
}}\right)  ^{1/2}, \label{eq_Rn}%
\end{equation}
where $n\in\mathbb{Z}$ is extended to integers by replacing $|n\rangle
\rightarrow|s_{n}n\rangle$. The corresponding LLs are analytically derived as%
\begin{align}
\varepsilon_{n}\left(  k_{x}^{\prime}\right)   &  =\hbar w_{x}^{\prime}%
k_{x}^{\prime}-\eta^{-2}s_{z}\gamma\hbar\upsilon_{B}k_{x}^{\prime}\delta
_{n,0}\nonumber\\
&  +\eta^{-3}s_{n}\hbar\sqrt{2|n|\omega_{c}^{2}+\eta^{2}\gamma^{2}\upsilon
_{B}^{2}k_{x}^{\prime2}}. \label{eq_mn}%
\end{align}
Detail derivation for the LLs and wavefunctions can be found in Appendix
\ref{DLW}.

As shown in Figs. \ref{Fig2}(a) and (b), the analytical expression agrees
perfectly with the numerical results. The Lorentz boost approach also has been
used to solve the LLs for a magnetic field applied along the symmetry axis of
the WSMs\cite{Serguei117}, similar to the case of an isotropic WSM subjected
to crossed electric and magnetic fields\cite{Deng99}. When $\alpha=0$, Eq.
(\ref{eq_mn}) recovers the LLs reported in Ref.\cite{Serguei117}. It is well
known that a Lorentz boost along the direction of the vector potential,
characterized by the Lorentz factor\cite{Deng99} $\eta=(1-\tanh^{2}%
\vartheta)^{-1/4}$, can eliminate the electric field or tilting term vertical
to the magnetic field and vector potential. After that, the LLs can be solved
analytically in the boosted reference frame\cite{Lukose98,Serguei117,Deng99}.
However, the Lorentz boost approach cannot yield an explicit expression for
the normalized wavefunction. More generally, we here derive both the LLs and
their corresponding wavefunction using the Bogoliubov transformation, where
the magnetic field can be applied in arbitrary directions. Notably, due to the
interaction with the magnetic field and the anisotropy of the Fermi velocity,
the vertical component of the tilt vector, i.e., perpendicular to the magnetic
field, can induce tilt of the LLs along the magnetic field, as implied in the
coupling term $\alpha\hbar w_{y}^{\prime}k_{x}^{\prime}$ within $\hbar
w_{x}^{\prime}k_{x}^{\prime}$ of Eq. (\ref{eq_mn}).

By comparing the Lorentz factor with Eq. (\ref{eq_RP}), we obtain
\begin{equation}
\tanh^{2}\vartheta=\frac{2\omega_{+}\omega_{-}}{\omega_{c}^{2}}=\frac
{w_{y}^{\prime2}}{\upsilon_{B}^{2}}+\frac{w_{z}^{2}}{\upsilon_{z}^{2}},
\end{equation}
which demonstrates that the effect of the vertical component of the tilt
vector is equivalent to that of a vertical electric field, or vice versa.
Since $\tanh^{2}\vartheta<1$, the Bogoliubov transformation, or equivalently
the Lorentz boost, is valid for
\begin{equation}
\frac{w_{y}^{\prime2}}{\upsilon_{B}^{2}}+\frac{w_{z}^{2}}{\upsilon_{z}^{2}%
}=t_{w}^{2}-\left(  \frac{w_{x}\upsilon_{y}^{2}\cos\varphi_{B}+w_{y}%
\upsilon_{x}^{2}\sin\varphi_{B}}{\upsilon_{x}\upsilon_{y}\upsilon_{B}}\right)
^{2}<1, \label{eq_tw2}%
\end{equation}
beyond which the LLs collapse. Specifically, the boundaries of the LL
collapse, as shown in Figs. \ref{Fig2}(d) and (e), are described by
\begin{equation}
(1-\frac{w_{z}^{2}}{\upsilon_{z}^{2}}-\frac{w_{x}^{2}}{\upsilon_{x}^{2}}%
)\tan^{2}\varphi_{B}+2\frac{w_{x}w_{y}}{\upsilon_{x}^{2}}\tan\varphi_{B}%
+\frac{\upsilon_{y}^{2}}{\upsilon_{x}^{2}}(1-\frac{w_{z}^{2}}{\upsilon_{z}%
^{2}}-\frac{w_{y}^{2}}{\upsilon_{y}^{2}})=0.
\end{equation}
This is consistent with the numerical results presented in Fig. \ref{Fig2}. In
type-\textrm{I} WSMs, since $t_{w}^{2}<1$, Eq. (\ref{eq_tw2}) is always
satisfied, ensuring that the LLs do not collapse. This is because the closed
Fermi surface allows the Weyl fermions to undergo complete cyclotron motion,
see the blue curve in Fig. \ref{Fig1}(b). In contrast, for type-\textrm{II}
WSMs where $t_{w}^{2}>1$, Eq. (\ref{eq_tw2}) can be violated. For example,
when $w_{x}=0$ and $w_{y}=0$, Eq. (\ref{eq_tw2}) is always violated and the
LLs must collapse. In this case, the cross-section of the Fermi surface
perpendicular to the magnetic field is always open, illustrated by the red
curves in Fig. \ref{Fig1}(c), which prevents the formation of LLs.

Interestingly, when $w_{x}$ or $w_{y}$ is finite, the LLs do not collapse even
in type-\textrm{II} WSMs for certain magnetic field directions, as illustrated
in Figs. \ref{Fig2}(d) and (e), where the boundaries of the phase transition
and the LL collapse do not coincide. This can be understood as follows. In
type-\textrm{II} WSMs, while the cross-section of the Fermi surfaces that
parallel to the tilt vector are open, see the red curves in Fig.
\ref{Fig1}(c), the cross-section vertical to the tilt vector remains closed,
as shown by the blue curves in Fig. \ref{Fig1}(c). For finite $w_{x}$ or
$w_{y}$, the tilt vector deviates from the $z$-axis and becomes nonorthogonal
to the $x$-$y$ plane. As a result, when the magnetic field is applied in the
$x$-$y$ plane, the cross-section of the Fermi surfaces can be closed in the
plane perpendicular to the magnetic field, as sketched in Fig. \ref{Fig1}(d),
which enables the complete cyclotron orbits necessary for the LL quantization.

For a fixed $k_{x}^{\prime}$, the cyclotron orbit is minimized when the
magnetic field is parallel to the tilt vector. According to the Sommerfeld's
quantization condition, the spacing between the LLs is maximal in this
configuration. As the angle between the magnetic field and the tilt vector
increases, the cyclotron orbits enlarge and transition from a circle to an
ellipse, causing the spacing between the LLs to decrease. This analysis is
consistent with the numerical results shown in Figs. \ref{Fig2}(a)-(c). The
underlying physics is also applicable to the case of type-\textrm{I} WSMs with
anisotropic Fermi velocities. \begin{figure}[ptb]
\centering \includegraphics[width=0.5\textwidth]{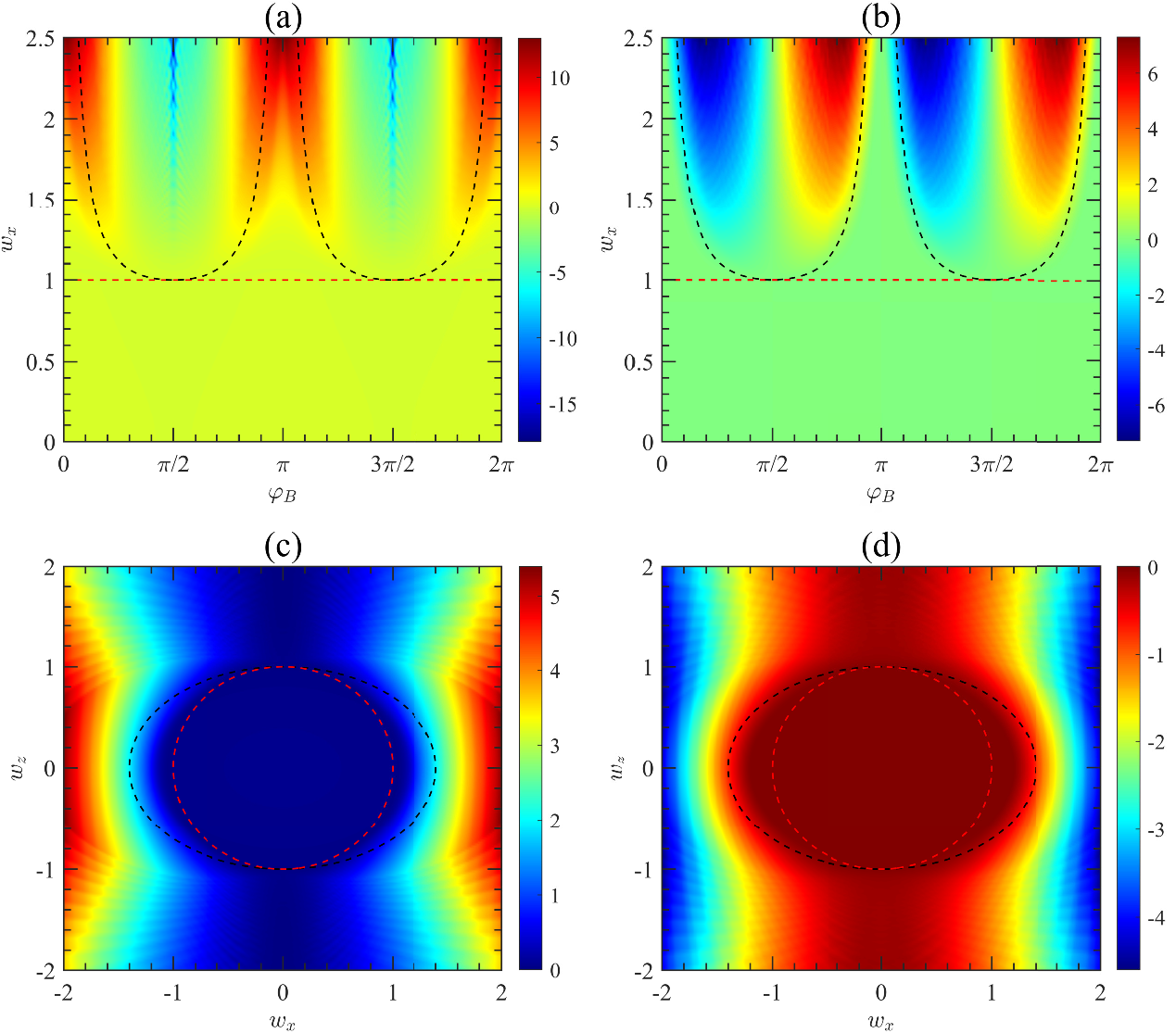}\caption{Distribution
of $\sigma_{\mathrm{L}}^{-}$ (left panel) and $\sigma_{\mathrm{T}}^{+}$ (right
panel) in the parameter spaces $w_{x} $-$\varphi_{B}$ (upper panel) and
$w_{x}$-$w_{z}$ (lower panel). The parameters are set as $E_{F}=0$,
$\Gamma=0.5\hbar\omega_{0}$, $\tau_{e}=5\hbar/\Gamma$, and other parameters
the same as Fig. \ref{Fig2}.}%
\label{Fig3}%
\end{figure}

\section{PHE signal for the LL collapse}

\label{PFC}

As discussed above, the LLs in anisotropic WSMs are highly sensitive to the
direction of the magnetic field, which suggests that the PHE can serve as a
powerful tool for detecting the LL collapse. Since the LLs do not collapse in
type-\textrm{I} WSMs but must collapse in type-\textrm{II} WSMs under an
appropriate magnetic field orientation, the magnetotransport signal from the
LL collapse can be used to determine the phase transition from type-\textrm{I}
to type-\textrm{II} WSMs. In the following, we apply an electric field in the
$x$-$y$ plane, denoted by $\boldsymbol{E}=E\left(  \cos\varphi_{E},\sin
\varphi_{E},0\right)  $, to investigate the PHE and extract the
magnetotransport signal associated with the LL collapse.

We calculate the conductivity tensor from the Kubo-Streda
formula\cite{Li99,Chen108}%
\begin{equation}
\sigma_{ij}=-\frac{\hbar e^{2}}{2\pi}\int_{-\infty}^{\infty}d\epsilon
\mathrm{Tr}[\hat{\upsilon}_{i}(\mathcal{G}^{<}\hat{\upsilon}_{j}%
\partial_{\epsilon}\mathcal{G}^{A}-\partial_{\epsilon}\mathcal{G}^{R}%
\hat{\upsilon}_{j}\mathcal{G}^{<})] \label{eq_XabG}%
\end{equation}
with $\mathcal{G}^{<}=(\mathcal{G}^{A}-\mathcal{G}^{R})f(\epsilon)$, where
$f\left(  \epsilon\right)  =1/[\exp(\frac{\epsilon-E_{F}}{k_{B}T})+1]$
represents the Fermi-Dirac distribution, and
\begin{equation}
\mathcal{G}^{R/A}=\frac{1}{\epsilon-\mathcal{\tilde{H}}_{\boldsymbol{k}}\pm
i\Gamma}=\sum_{n}\frac{|\Psi_{n}\rangle\langle\Psi_{n}|}{\epsilon
-\varepsilon_{n}\left(  k_{x}^{\prime}\right)  \pm i\Gamma} \label{eq_GRA2}%
\end{equation}
denotes the unperturbed retarded/advanced Green's function. In the
representation of $\tilde{\sigma}_{x}$, the velocity operator becomes%
\begin{equation}
\hat{\upsilon}_{i}=\frac{1}{i\hbar}[r_{i},\mathcal{\tilde{H}}_{\boldsymbol{k}%
}]=\frac{\partial\mathcal{\tilde{H}}_{\boldsymbol{k}}}{\hbar\partial k_{i}}.
\end{equation}
Upon application of the electric and magnetic fields, the chemical potential
in each Weyl valley is modified by $\Delta\mu_{\chi}=-eE_{j}\langle
\hat{\upsilon}_{j}\rangle_{\chi}\tau_{e}$, where $\tau_{e}$ is the intervalley
relaxation time and the average $\langle\cdots\rangle_{\chi}$ runs over all
electron states at the Fermi level in the $\chi$ valley\cite{Deng99,Deng122}.

By substituting Eq. (\ref{eq_GRA2}) into Eq. (\ref{eq_XabG}), combined with
the chiral anomaly of WSMs, we can decompose the conductivity tensor into
three terms, i.e., $\sigma_{ij}=\sigma_{ij}^{\mathrm{D}}+\sigma_{ij}%
^{\mathrm{inter}}+\Delta\sigma_{ij}^{\mathrm{ch}}$, where%
\begin{equation}
\sigma_{ij}^{\mathrm{D}}=-\frac{\hbar e^{2}}{2\pi\ell_{B}^{2}}\frac{1}{\Gamma
}\sum_{k_{x}^{\prime},n}\frac{\partial f_{n}\left(  k_{x}^{\prime}\right)
}{\partial\varepsilon_{n}\left(  k_{x}^{\prime}\right)  }\langle\Psi_{n}%
|\hat{\upsilon}_{i}|\Psi_{n}\rangle\langle\Psi_{n}|\hat{\upsilon}_{j}|\Psi
_{n}\rangle\label{eq_DR}%
\end{equation}
is the Drude conductivity from intraband transitions,%
\begin{equation}
\sigma_{ij}^{\mathrm{inter}}=-\frac{e^{2}}{\hbar}\sum_{k_{x}^{\prime},m\neq
n}\frac{f_{n}\left(  k_{x}^{\prime}\right)  \omega_{mn}^{2}}{2\pi\ell_{B}^{2}%
}\frac{\Omega_{ij}^{mn}\cos2\vartheta_{mn}-\mathcal{R}_{ij}^{mn}\sin
2\vartheta_{mn}}{\omega_{mn}^{2}+\Gamma^{2}/\hbar^{2}} \label{eq_LHall}%
\end{equation}
represents the interband contribution due to the Berry curvature and quantum
metric, and
\begin{equation}
\Delta\sigma_{ij}^{\mathrm{ch}}=-\frac{\hbar e^{2}}{2\pi\ell_{B}^{2}}%
\sum_{k_{x}^{\prime},n}\frac{\partial f_{n}\left(  k_{x}^{\prime}\right)
}{\partial\varepsilon_{n}\left(  k_{x}^{\prime}\right)  }\langle\Psi_{n}%
|\hat{\upsilon}_{i}|\Psi_{n}\rangle\langle\hat{\upsilon}_{j}\rangle_{\chi}%
\tau_{e}%
\end{equation}
accounts for the magnetoconctivity originating from the chiral
anomaly\cite{Deng99,Deng122}. Here, we noted $f_{n}\left(  k_{x}^{\prime
}\right)  =f\left(  \epsilon\right)  |_{\epsilon\rightarrow\varepsilon
_{n}\left(  k_{x}^{\prime}\right)  }$, $\vartheta_{mn}=\cot^{-1}\left(
\hbar\omega_{mn}/\Gamma\right)  $, and $\hbar\omega_{mn}=\varepsilon
_{m}\left(  k_{x}^{\prime}\right)  -\varepsilon_{n}\left(  k_{x}^{\prime
}\right)  $. The Berry curvature and quantum metric, defined respectively as%
\begin{align}
\Omega_{ij}^{mn}  &  =\frac{2\operatorname{Im}\langle\Psi_{m}|\hat{\upsilon
}_{i}|\Psi_{n}\rangle\langle\Psi_{n}|\hat{\upsilon}_{j}|\Psi_{m}\rangle
}{\left(  \varepsilon_{m}-\varepsilon_{n}\right)  ^{2}}\hbar^{2},\\
\mathcal{R}_{ij}^{mn}  &  =\frac{2\operatorname{Re}\langle\Psi_{m}%
|\hat{\upsilon}_{i}|\Psi_{n}\rangle\langle\Psi_{n}|\hat{\upsilon}_{j}|\Psi
_{m}\rangle}{\left(  \varepsilon_{m}-\varepsilon_{n}\right)  ^{2}}\hbar^{2},
\end{align}
are induced by the transitions between the LLs $\varepsilon_{m}\left(
k_{x}^{\prime}\right)  $ and $\varepsilon_{n}\left(  k_{x}^{\prime}\right)  $.
For $\hbar\omega_{mn}\gg\Gamma$, $\cos\vartheta_{mn}=1$ and $\sin
\vartheta_{mn}=\Gamma/\left(  \hbar\omega_{mn}\right)  $. The Drude, Berry
curvature and quantum metric contributions can be distinguished by their
dependence on the intravalley relaxation time $\sim\hbar/\Gamma$, with the
Drude contribution $\propto\Gamma^{-1}$, the Berry curvature contribution
$\propto\Gamma^{0}$ and the quantum metric contribution $\propto\Gamma$.
\begin{figure}[ptb]
\centering \includegraphics[width=0.495\textwidth]{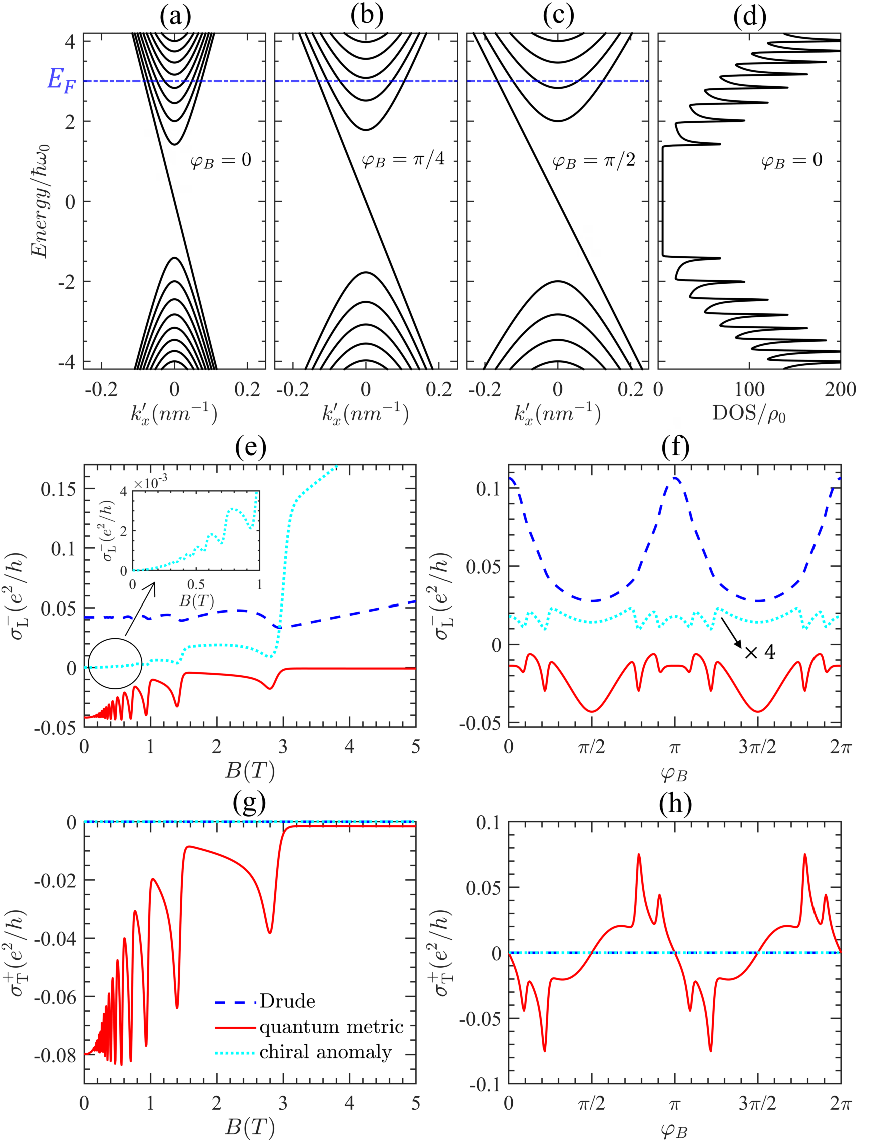}\caption{ (a)-(c)
The LLs for $B=1$ T and varied $\varphi_{B}$, where the blue dash-dotted lines
indicate the Fermi level. (d) Quantum oscillation of the DOSs with respect to
energy, in which $\rho_{0}=1/(2\pi\ell_{B}^{2})$. (e)-(f) $\sigma_{L}^{-}$ and
(g)-(h) $\sigma_{T}^{+}$ vs the magnitude and direction of the magnetic field
for $\varphi_{B}=\pi/4$ in (e) and (g), and $B=1$ T in (f) and (h). Here, we
set $E_{F}=3\hbar\omega_{0}|_{B=1\mathrm{T}}$, $\boldsymbol{w} = 0$, and the
rest parameters the same as Fig. \ref{Fig3}.}%
\label{Fig4}%
\end{figure}

For comparison with the previous results\cite{Taskin8,Weng106}, we denote
$x^{\prime}$ ($y^{\prime}$) as $\Vert$ ($\perp$), corresponding to the
direction parallel (vertical) to the magnetic field. According to Eq.
(\ref{eq_XabG}), we can derive the longitudinal conductivity $\sigma
_{\mathrm{L}} $ and the planar Hall conductivity $\sigma_{\mathrm{PHE}}$ as%
\begin{align}
\sigma_{\mathrm{L}}  &  =\sigma_{\perp}+\Delta\sigma\cos^{2}\varphi
-(\sigma_{\perp\parallel}+\sigma_{\parallel\perp})\sin\varphi\cos
\varphi,\label{eq_SL}\\
\sigma_{\mathrm{PHE}}  &  =\sigma_{\parallel\perp}+\Delta\sigma\sin\varphi
\cos\varphi-(\sigma_{\perp\parallel}+\sigma_{\parallel\perp})\sin^{2}\varphi,
\label{eq_SP}%
\end{align}
where $\Delta\sigma=\sigma_{\parallel}-\sigma_{\perp}$ and $\varphi
=\varphi_{B}-\varphi_{E}$ is the angle between the electric and magnetic
fields. Here, $\sigma_{\parallel}=\sigma_{x^{\prime}x^{\prime}}$ and
$\sigma_{\perp}=\sigma_{y^{\prime}y^{\prime}}$ represent the longitudinal
conductivities when the magnetic field is parallel and orthogonal to the
electric field, respectively. If $\sigma_{\parallel\perp}$ and $\sigma
_{\perp\parallel}$ are negligible, Eqs. (\ref{eq_SL}) and (\ref{eq_SP}) reduce
to the common relation for the PHE\cite{Taskin8}. For convenience, we rewrite
them in a more symmetric form%
\begin{align}
\sigma_{\mathrm{L}}  &  =\sigma_{\mathrm{L}}^{+}+\sigma_{\mathrm{L}}^{-}%
\cos2\varphi-\sigma_{\mathrm{T}}^{+}\sin2\varphi,\label{eq_LS}\\
\sigma_{\mathrm{PHE}}  &  =\sigma_{\mathrm{T}}^{-}+\sigma_{\mathrm{L}}^{-}%
\sin2\varphi+\sigma_{\mathrm{T}}^{+}\cos2\varphi, \label{eq_TS}%
\end{align}
where $\sigma_{\mathrm{L}}^{\pm}=(\sigma_{\parallel}\pm\sigma_{\perp})/2$ and
$\sigma_{\mathrm{T}}^{\pm}=(\sigma_{\perp\parallel}\pm\sigma_{\parallel\perp
})/2$. It is apparent that $\sigma_{\mathrm{T}}^{+}$ and $\sigma_{\mathrm{T}%
}^{-}$ correspond respectively to the symmetric and antisymmetric parts of the
in-plane transverse conductivity, while $\sigma_{\mathrm{L}}^{-} $ reflects
the anisotropic longitudinal magnetoresistance. The PHE is observable for
finite $\sigma_{\mathrm{L}}^{-}$ or/and $\sigma_{\mathrm{T}}^{+}$. Since
$\Omega_{ij}^{mn}=-\Omega_{ji}^{mn}$ is antisymmetric under interchange of the
subscripts, the Berry curvature does not contribute to the symmetric part of
the PHE. On the contrary, the quantum metric is symmetric when interchanging
the subscripts and therefore contributes to the symmetric part of the
interband PHE. \begin{figure}[ptb]
\centering \includegraphics[width=0.49\textwidth]{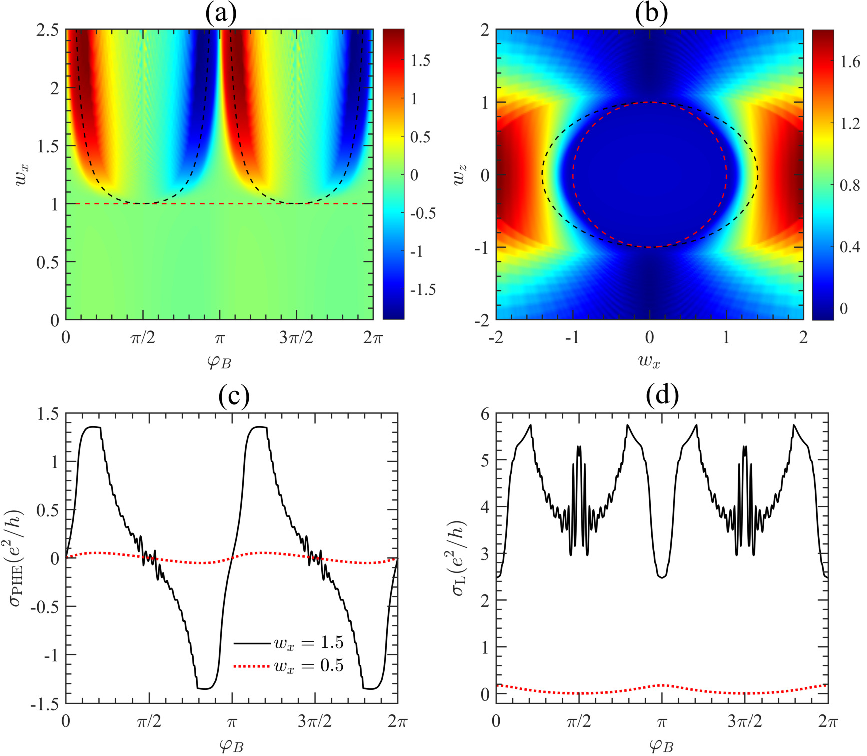}\caption{
Distribution of the planar Hall conductivity $\sigma_{\mathrm{PHE}}$ in the
parameter spaces (a) $w_{x}$-$\varphi_{B}$ and (b) $w_{x}$-$w_{z}$. (c) The
planar Hall conductivity and (d) the longitudinal conductivity as function of
$\varphi_{B}$ for $w_{z}=0$ and varied $w_{x}$. Other parameters are the same as in Fig. \ref{Fig3}. }%
\label{Fig5}%
\end{figure}

In Fig. \ref{Fig3}, we present the numerical results for $\sigma_{\mathrm{L}%
}^{-}$ and $\sigma_{\mathrm{T}}^{+}$, and $\sigma_{\mathrm{T}}^{-}$ is
vanishing in our case. As shown in Figs. \ref{Fig3}(a) and (b), in the strong
magnetic field regime, i.e., $E_{F}\ll\sqrt{2}\hbar\omega_{c}$, $\sigma
_{\mathrm{L}}^{-}$ and $\sigma_{\mathrm{T}}^{+}$ are nearly independent of
$\varphi_{B}$ for type-\textrm{I} WSMs, but are enhanced significantly for
type-\textrm{II} WSMs around the boundary of the LL collapse. Moreover,
$\sigma_{\mathrm{L}}^{-}$ and $\sigma_{\mathrm{T}}^{+}$ change suddenly at the
boundary of the LL collapse when tuning the tilt vector, see Figs.
\ref{Fig3}(c) and (d). As a result, it is expected that $\sigma_{\mathrm{L}}$
and $\sigma_{\mathrm{PHE}}$ will exhibit a complex angular dependence beyond
the explicit $\cos2\varphi$ and $\sin2\varphi$ in Eqs. (\ref{eq_LS}) and
(\ref{eq_TS}).

In the presence of anisotropy in the Fermi velocity, the cyclotron orbits
deform as $\varphi_{B}$ varies, so that the spacing of the LLs in
type-\textrm{I} WSMs also depends on the angle of the magnetic field, as
depicted in Figs. \ref{Fig4}(a)-(c). Consequently, the Fermi level for
relatively weak magnetic fields, i.e., $E_{F}>\sqrt{2}\hbar\omega_{c}$, can
intersect the Van Hove singularities of the LLs with rotating the magnetic
field. The density of states (DOSs) exhibits sharp peaks when the Fermi level
crosses the Van Hove singularities, as illustrated in Fig. \ref{Fig4}(d). As a
result, $\sigma_{\mathrm{L}}^{-}$ and $\sigma_{\mathrm{T}}^{+}$ display
quantum oscillations when varying either the magnitude or direction of the
magnetic field, as shown in Figs. \ref{Fig4}(e)-(h). It is well known that in
WSMs, when the magnetic and electric fields are aligned, the chiral anomaly
will be activated, resulting in the positive longitudinal
magnetoconductivity\cite{Deng99,Deng122}. The chiral anomaly mainly
contributes to $\sigma_{\parallel}$ and is turned off for $\sigma_{\perp}$,
such that $\sigma_{\mathrm{L}}^{-}\neq0$ and thereby inducing the PHE.
Interestingly, we find that the quantum metric can also contribute to the
positive longitudinal magnetoconductivity, not only for $\sigma_{\mathrm{L}%
}^{-}$, but also for $\sigma_{\mathrm{T}}^{+}$. The quantum metric
contribution is attributed to the impurity-scattering-induced transitions
between different LLs, determined by the relative magnitude of $\Gamma$ and
the spacing between the LLs. As shown in Fig. \ref{Fig4}(e), in the weak
magnetic field regime, the contribution from the quantum metric dominates over
the chiral anomaly. With increasing the magnitude of the magnetic field, the
contribution from the quantum metric gradually decreases, and the chiral
anomaly becomes dominant in the strong magnetic field regime. Both the chiral
anomaly and quantum metric contributions are sensitive to the direction of the
magnetic field, as shown in Fig. \ref{Fig4}(f). As a result, $\sigma
_{\mathrm{L}}$ and $\sigma_{\mathrm{PHE}}$ exhibit complex angular
dependences, superimposed by quantum oscillations, as shown in Fig. \ref{Fig5}.

During the phase transition from type-\textrm{I} to type-\textrm{II} WSMs, the
number of transport channels around the Weyl nodes increases significantly.
Therefore, it is expected that the transport coefficients will change
drastically as the tilt vector varies. The distribution of the planar Hall
conductivity in the parameter spaces $w_{x}$-$\varphi_{B}$ and $w_{x}$-$w_{z}$
is shown in Figs. \ref{Fig5} (a) and (b), respectively. Clearly, the PHE is
enhanced around the boundaries of the LL collapse, making the amplitude of the
PHE increase significantly in type-\textrm{II} WSMs. Due to the LL collapse,
the longitudinal and planar Hall conductivities in type-\textrm{II} WSMs
deviate heavily from the normal angular dependence, decorated with quantum
oscillations, as illustrated in Figs. \ref{Fig5} (c) and (d). When the
magnetic field is rotated in the plane vertical to the tilt vector, the
boundaries of the LL collapse and the phase transition coincide. Therefore,
the irregular angular-dependent planar Hall conductivity superimposed with
quantum oscillations provides a strong signal for the LL collapse and can
serve as an experimental signature for the phase transition from
type-\textrm{I} to type-\textrm{II} WSMs.

\section{Conclusions}

In summary, we have investigated the PHE in anisotropic WSMs under strong
magnetic fields. The LLs and wavefunctions are analytically derived using the
Bogoliubov transformation. We demonstrated that the LL collapse is closely
related to the phase transition of WSMs, i.e., the LLs do not collapse in
type-\textrm{I} WSMs but must collapse in type-\textrm{II} WSMs when varying
the orientation of the magnetic field. The magnetotransport signal of the LL
collapse, which manifests as significant enhancement and quantum oscillations
simultaneously in the longitudinal and planar Hall conductivities, can be used
to detect the phase transition from type-\textrm{I} to type-\textrm{II} WSMs.

\section{Acknowledgement}

This work was supported by the National NSF of China under Grants No.
12274146, No. 12174121 and No. 12104167; the Guang dong Basic and Applied
Basic Research Foundation under Grant No. 2023B1515020050; the Guang dong NSF
of China under Grant No. 2024A1515011300; and the Guangdong Provincial Quantum
Science Strategic Initiative under Grant No. GDZX2401002.

\bibliography{bibcfy2025}
\appendix
\section{Derivation for the form of the Bogoliubov transformation}
\label{DBT}
The wavefunction of Hamiltonian (\ref{eq_Hp}) is a two-component spinor
$\Psi=(|\psi\rangle,|\varphi\rangle)^{T}$, which satisfies the simultaneous
equations%
\begin{align}
\sqrt{2}\omega_{c}\hat{\Pi}|\varphi\rangle &  =\left(  \omega_{\varepsilon
}-\gamma\upsilon_{B}k_{x}^{\prime}-\omega_{-}\hat{\Pi}-\omega_{+}\hat{\Pi
}^{\dag}\right)  |\psi\rangle,\label{eq_A1}\\
\sqrt{2}\omega_{c}\hat{\Pi}^{\dag}|\psi\rangle &  =\left(  \omega
_{\varepsilon}+\gamma\upsilon_{B}k_{x}^{\prime}-\omega_{-}\hat{\Pi}-\omega
_{+}\hat{\Pi}^{\dag}\right)  |\varphi\rangle, \label{eq_A2}%
\end{align}
where $\hbar\omega_{\varepsilon}=\varepsilon-\hbar w_{x}^{\prime}k_{x}%
^{\prime}$ and $\varepsilon$ represents the eigenenergy. Applying the operator
$\sqrt{2}\omega_{c}\hat{\Pi}^{\dag}$ to Eq. (\ref{eq_A1}) and applying the
operator $\sqrt{2}\omega_{c}\hat{\Pi}$ to Eq. (\ref{eq_A2}), we can arrive at
the matrix equation%
\begin{equation}
\left(
\begin{array}
[c]{cc}%
\mathcal{\hat{F}}-s_{z}\omega_{c}^{2} & -\sqrt{2}s_{z}\omega_{c}\omega_{+}\\
\sqrt{2}s_{z}\omega_{c}\omega_{-} & \mathcal{\hat{F}}+s_{z}\omega_{c}^{2}%
\end{array}
\right)  \left(
\begin{array}
[c]{c}%
|\psi\rangle\\
|\varphi\rangle
\end{array}
\right)  =0, \label{eq_AW}%
\end{equation}
where
\begin{align}
\mathcal{\hat{F}}  &  =\omega_{\varepsilon}^{2}-\gamma^{2}\upsilon_{B}%
^{2}k_{x}^{\prime2}-\left(  \omega_{c}^{2}-\omega_{+}\omega_{-}\right)
\left(  2\hat{\Pi}^{\dag}\hat{\Pi}+s_{z}\right) \nonumber\\
&  +\omega_{-}^{2}\hat{\Pi}\hat{\Pi}+\omega_{+}^{2}\hat{\Pi}^{\dag}\hat{\Pi
}^{\dag}-2\omega_{\varepsilon}\left(  \omega_{-}\hat{\Pi}+\omega_{+}\hat{\Pi
}^{\dag}\right)  . \label{eq_AF}%
\end{align}

To eliminate the terms related to $\hat{\Pi}\hat{\Pi}$ and $\hat{\Pi}^{\dag
}\hat{\Pi}^{\dag}$, we perform the Bogoliubov transformation%
\begin{align}
\hat{\Pi}  &  =p\hat{\beta}+q^{\ast}\hat{\beta}^{\dag},\label{eq_AC}\\
\hat{\Pi}^{\dag}  &  =q\hat{\beta}+p^{\ast}\hat{\beta}^{\dag}. \label{eq_ACD}%
\end{align}
For simplicity, it is assumed that $p$ is real. By substituting Eqs.
(\ref{eq_AC}) and (\ref{eq_ACD}) into Eq. (\ref{eq_AF}), and setting
\begin{align}
2\left(  \omega_{c}^{2}-\omega_{+}\omega_{-}\right)  qp-\omega_{+}%
^{2}qq-\omega_{-}^{2}pp  &  =0,\label{eq_p}\\
2\left(  \omega_{c}^{2}-\omega_{+}\omega_{-}\right)  p^{\ast}q^{\ast}%
-\omega_{+}^{2}p^{\ast}p^{\ast}-\omega_{-}^{2}q^{\ast}q^{\ast}  &  =0,
\label{eq_q}%
\end{align}
we can arrive at
\begin{align}
\mathcal{\hat{F}}  &  =\omega_{\varepsilon}^{2}-\gamma^{2}\upsilon_{B}%
^{2}k_{x}^{\prime2}-s_{z}\left(  \omega_{c}^{2}-\omega_{+}\omega_{-}\right)
\nonumber\\
&  -2\omega_{\varepsilon}\left[  \left(  \omega_{-}p+\omega_{+}q\right)
\hat{\beta}+\left(  \omega_{-}q^{\ast}+\omega_{+}p^{\ast}\right)  \hat{\beta
}^{\dag}\right] \nonumber\\
&  -\left[  2\left(  \omega_{c}^{2}-\omega_{+}\omega_{-}\right)  pp^{\ast
}-\omega_{+}^{2}p^{\ast}q-\omega_{-}^{2}q^{\ast}p\right]  \hat{\beta}^{\dag
}\hat{\beta}\nonumber\\
&  -\left[  2\left(  \omega_{c}^{2}-\omega_{+}\omega_{-}\right)  qq^{\ast
}-\omega_{+}^{2}qp^{\ast}-\omega_{-}^{2}pq^{\ast}\right]  \hat{\beta}%
\hat{\beta}^{\dag}. \label{eq_AFB1}%
\end{align}
From Eqs. (\ref{eq_p}) and (\ref{eq_q}), together with the commutation
relation
\begin{equation}
\left[  \hat{\Pi},\hat{\Pi}^{\dag}\right]  =\left(  pp^{\ast}-qq^{\ast
}\right)  \left[  \hat{\beta},\hat{\beta}^{\dag}\right]  =s_{z},
\end{equation}
we can solve for%
\begin{align}
pp^{\ast}  &  =\lambda_{+}^{2},\label{eq_pp}\\
qq^{\ast}  &  =\lambda_{-}^{2},\label{eq_qq}\\
pq^{\ast}  &  =\lambda_{+}\lambda_{-}\frac{\omega_{+}}{\omega_{-}},
\label{eq_pq}%
\end{align}
where $\left[  \hat{\beta},\hat{\beta}^{\dag}\right]  =1$, $\lambda_{\pm
}=\frac{\eta\pm s_{z}\eta^{-1}}{2}$, and
\begin{equation}
\eta=\frac{1}{\left(  1-2\frac{\omega_{+}\omega_{-}}{\omega_{c}^{2}}\right)
^{1/4}}.
\end{equation}
Accordingly, $p=\lambda_{+}$ and $q=\lambda_{-}\frac{\omega_{-}}{\omega_{+}}$.
Substituting Eqs. (\ref{eq_pp})-(\ref{eq_pq}) into Eq. (\ref{eq_AFB1}) yields%
\begin{align}
\mathcal{\hat{F}}  &  =\omega_{\varepsilon}^{2}-\gamma^{2}\upsilon_{B}%
^{2}k_{x}^{\prime2}-\eta^{-2}\omega_{c}^{2}\left(  2\hat{\beta}^{\dag}%
\hat{\beta}+1\right) \nonumber\\
&  -2\omega_{\varepsilon}\left[  \left(  \omega_{-}p+\omega_{+}q\right)
\hat{\beta}+\left(  \omega_{-}q^{\ast}+\omega_{+}p^{\ast}\right)  \hat{\beta
}^{\dag}\right]  . \label{eq_AFB2}%
\end{align}

To further eliminate the terms associated with $\hat{\beta}$ and $\hat{\beta
}^{\dag}$ in Eq. (\ref{eq_AFB2}), we replace%
\begin{align}
\hat{\beta}  &  \rightarrow\hat{\beta}+K,\\
\hat{\beta}^{\dag}  &  \rightarrow\hat{\beta}^{\dag}+K^{\ast},
\end{align}
and set the coefficient in front of $\hat{\beta}$ and $\hat{\beta}^{\dag}$ to
be zero. Subsequently, we can obtain
\begin{equation}
K=-\eta^{2}\frac{\omega_{\varepsilon}}{\omega_{c}^{2}}\left(  \omega
_{-}q^{\ast}+\omega_{+}p^{\ast}\right)
\end{equation}
and simplify $\mathcal{\hat{F}}$ as%
\begin{equation}
\mathcal{\hat{F}}=\eta^{4}\omega_{\varepsilon}^{2}-\gamma^{2}\upsilon_{B}%
^{2}k_{x}^{\prime2}-\eta^{-2}\omega_{c}^{2}\left(  2\hat{\beta}^{\dag}%
\hat{\beta}+1\right)  ,
\end{equation}
as displayed in Eq. (\ref{eq_MF}) of the main texts. Consequently, the
Bogoliubov transformation becomes%
\begin{align}
\hat{\Pi}  &  =\lambda_{+}\hat{\beta}+\lambda_{-}\frac{\omega_{+}}{\omega_{-}%
}\hat{\beta}^{\dag}-2\eta^{2}\lambda_{+}\lambda_{-}\frac{\omega_{\varepsilon}%
}{\omega_{-}},\label{eq_CA1}\\
\hat{\Pi}^{\dag}  &  =\lambda_{+}\hat{\beta}^{\dag}+\lambda_{-}\frac
{\omega_{-}}{\omega_{+}}\hat{\beta}-2\eta^{2}\lambda_{+}\lambda_{-}%
\frac{\omega_{\varepsilon}}{\omega_{+}}, \label{eq_CA2}%
\end{align}
where Eq. (\ref{eq_CA1}) is exactly Eq. (\ref{eq_Pi}) in the main texts.

\section{Derivation for the form of the wavefunction}
\label{DFW}
The components of the wavefunction in Eq. (\ref{eq_AW}) can be expanded in
terms of the eigenstates of $\hat{\beta}^{\dag}\hat{\beta}$ as
\begin{align}
|\psi\rangle &  =\sum_{n}C_{n}|n\rangle,\label{eq_phi}\\
|\varphi\rangle &  =\sum_{m}D_{m}|m\rangle, \label{eq_psi}%
\end{align}
where $\hat{\beta}^{\dag}\hat{\beta}|n\rangle=n|n\rangle$, and
\begin{align}
\hat{\beta}|n\rangle &  =\sqrt{n}|n-1\rangle,\\
\hat{\beta}^{\dag}|n\rangle &  =\sqrt{n+1}|n+1\rangle.
\end{align}
By substituting the above expansion into Eq. (\ref{eq_AW}) , we can obtain the
equations%
\begin{align}
\left(  \mathcal{\hat{F}}-s_{z}\omega_{c}^{2}\right)  \sum_{n}C_{n}%
|n\rangle-\sqrt{2}s_{z}\omega_{c}\omega_{+}\sum_{m}D_{m}|m\rangle &
=0,\label{eq_FB1}\\
\sqrt{2}s_{z}\omega_{c}\omega_{-}\sum_{n}C_{n}|n\rangle+\left(  \mathcal{\hat
{F}}+s_{z}\omega_{c}^{2}\right)  \sum_{m}D_{m}|m\rangle &  =0. \label{eq_FB2}%
\end{align}
Then, we multiply both sides of Eqs. (\ref{eq_FB1}) and (\ref{eq_FB2}) by
$\langle l|$ and derive for%
\begin{equation}
\left(
\begin{array}
[c]{cc}%
\mathcal{F}_{l}-s_{z}\omega_{c}^{2} & -\sqrt{2}s_{z}\omega_{c}\omega_{+}\\
\sqrt{2}s_{z}\omega_{c}\omega_{-} & \mathcal{F}_{l}+s_{z}\omega_{c}^{2}%
\end{array}
\right)  \left(
\begin{array}
[c]{c}%
C_{l}\\
D_{l}%
\end{array}
\right)  =0, \label{eq_Amn}%
\end{equation}
where
\begin{equation}
\mathcal{F}_{l}=\eta^{4}\omega_{\varepsilon}^{2}-\eta^{-2}\omega_{c}%
^{2}\left(  2l+1\right)  -\gamma^{2}\upsilon_{B}^{2}k_{x}^{\prime2},
\end{equation}
and the orthonormality $\langle m|n\rangle=\delta_{mn}$ has been adopted.

By solving the secular equation (\ref{eq_Amn}), we can obtain
\begin{equation}
\eta^{4}\omega_{\varepsilon}^{2}=\gamma^{2}\upsilon_{B}^{2}k_{x}^{\prime
2}+\eta^{-2}\omega_{c}^{2}\left(  2l+1\right)  \pm\eta^{-2}\omega_{c}^{2}
\label{eq_AL}%
\end{equation}
and%
\begin{equation}
|\Phi_{l}\rangle=\left(
\begin{array}
[c]{c}%
C_{l}\\
D_{l}%
\end{array}
\right)  =\left(
\begin{array}
[c]{c}%
\sqrt{2}s_{z}\omega_{c}\omega_{+}\\
\mathcal{F}_{l}-s_{z}\omega_{c}^{2}%
\end{array}
\right)  .
\end{equation}
As a result, for a fixed $\omega_{\varepsilon}$, there are two solutions to
Eq. (\ref{eq_AL}), namely,%
\begin{align}
\eta^{4}\omega_{\varepsilon}^{2}  &  =\gamma^{2}\upsilon_{B}^{2}k_{x}%
^{\prime2}+\eta^{-2}\omega_{c}^{2}\left(  2n+1\right)  -\eta^{-2}\omega
_{c}^{2},\\
\eta^{4}\omega_{\varepsilon}^{2}  &  =\gamma^{2}\upsilon_{B}^{2}k_{x}%
^{\prime2}+\eta^{-2}\omega_{c}^{2}\left(  2m+1\right)  +\eta^{-2}\omega
_{c}^{2},
\end{align}
from which we can determine
\begin{equation}
n-m=1.
\end{equation}
Therefore, the general solution of the wavefunction is a linear superposition
of $|\Phi_{n}\rangle$ and $|\Phi_{n-1}\rangle$, i.e.,
\begin{align}
|\Psi_{n}\rangle &  =\left(  |\psi_{n}\rangle,|\varphi_{n}\rangle\right)
^{T}\nonumber\\
&  =C_{n}\left(
\begin{array}
[c]{c}%
\sqrt{\lambda_{+}}\\
-\frac{\omega_{-}}{\sqrt{\omega_{+}\omega_{-}}}\sqrt{\lambda_{-}}%
\end{array}
\right)  |n-1\rangle+D_{n}\left(
\begin{array}
[c]{c}%
-\frac{\omega_{+}}{\sqrt{\omega_{+}\omega_{-}}}\sqrt{\lambda_{-}}\\
\sqrt{\lambda_{+}}%
\end{array}
\right) |n\rangle\nonumber\\
&  =\sqrt{\lambda_{+}}\left(
\begin{array}
[c]{c}%
C_{n}|n-1\rangle\\
D_{n}|n\rangle
\end{array}
\right)  -\sqrt{\frac{\lambda_{-}}{\omega_{+}\omega_{-}}}\left(
\begin{array}
[c]{c}%
\omega_{+}D_{n}|n\rangle\\
\omega_{-}C_{n}|n-1\rangle
\end{array}
\right)  , \label{eq_APsi}%
\end{align}
which is Eq. (\ref{eq_WF}) of the main texts.

\section{Derivation for the LLs and their wavefunction}

\label{DLW}

According to Eqs. (\ref{eq_CA1})-(\ref{eq_CA2}) and (\ref{eq_APsi}), we can
derive%
\begin{equation}
|\psi_{n}\rangle=\lambda_{+}C_{n}|n-1\rangle-\frac{\eta\omega_{+}}{\sqrt
{2}\omega_{c}}D_{n}|n\rangle,\label{eq_C1}%
\end{equation}%
\begin{equation}
\hat{\Pi}|\varphi_{n}\rangle=\left(
\begin{array}
[c]{c}%
-\lambda_{+}\frac{\eta\omega_{-}}{\sqrt{2}\omega_{c}}C_{n}\sqrt{n-1}%
|n-2\rangle\\
+\lambda_{+}\left(  \lambda_{+}D_{n}\sqrt{n}+\sqrt{2}\frac{\eta^{3}%
\omega_{\varepsilon}}{\omega_{c}}\lambda_{-}C_{n}\right)  |n-1\rangle\\
-\lambda_{-}\left(  2\frac{\eta^{2}\omega_{\varepsilon}}{\omega_{-}}%
\lambda_{+}^{2}D_{n}+\frac{\eta\omega_{+}}{\sqrt{2}\omega_{c}}C_{n}\sqrt
{n}\right)  |n\rangle\\
+\frac{\omega_{+}}{\omega_{-}}\lambda_{+}\lambda_{-}D_{n}\sqrt{n+1}|n+1\rangle
\end{array}
\right)  ,\label{eq_C2}%
\end{equation}%
\begin{equation}
\left(  \omega_{-}\hat{\Pi}+\omega_{+}\hat{\Pi}^{\dag}\right)  |\psi
_{n}\rangle=\left(
\begin{array}
[c]{c}%
\eta\omega_{-}\lambda_{+}C_{n}\sqrt{n-1}|n-2\rangle\\
-\eta^{2}\left(  4\omega_{\varepsilon}\lambda_{+}^{2}\lambda_{-}C_{n}%
+\frac{\omega_{+}\omega_{-}}{\sqrt{2}\omega_{c}}D_{n}\sqrt{n}\right)
|n-1\rangle\\
+\eta\lambda_{+}\omega_{+}\left(  C_{n}\sqrt{n}+4\lambda_{-}\frac{\eta
^{2}\omega_{\varepsilon}}{\sqrt{2}\omega_{c}}D_{n}\right)  |n\rangle\\
-\frac{\eta^{2}\omega_{+}^{2}}{\sqrt{2}\omega_{c}}D_{n}\sqrt{n+1}|n+1\rangle
\end{array}
\right)  .\label{eq_C3}%
\end{equation}
For $n>0$, by substituting Eqs. (\ref{eq_C1})-(\ref{eq_C3}) into Eq.
(\ref{eq_A1}), we can obtain%
\begin{equation}
\left(
\begin{array}
[c]{cc}%
\eta^{2}\omega_{\varepsilon}-s_{z}\gamma\upsilon_{B}k_{x}^{\prime} &
-\eta^{-1}\sqrt{2n}\omega_{c}\\
-\eta^{-1}\sqrt{2n}\omega_{c} & \eta^{2}\omega_{\varepsilon}+s_{z}%
\gamma\upsilon_{B}k_{x}^{\prime}%
\end{array}
\right)  \left(
\begin{array}
[c]{c}%
C_{n}\\
D_{n}%
\end{array}
\right)  =0,
\end{equation}
from which we can determine%
\begin{equation}
\omega_{\varepsilon}=\pm\eta^{-3}\sqrt{2n\omega_{c}^{2}+\eta^{2}\gamma
^{2}\upsilon_{B}^{2}k_{x}^{\prime2}},
\end{equation}%
\begin{align}
\omega_{\varepsilon,+} &  =+\eta^{-3}\sqrt{2n\omega_{c}^{2}+\eta^{2}\gamma
^{2}\upsilon_{B}^{2}k_{x}^{\prime2}},\\
|\Phi_{n,+}\rangle &  =\frac{1}{\sqrt{2}}\left(
\begin{array}
[c]{c}%
\sqrt{\eta^{-1}+\frac{s_{z}\gamma\upsilon_{B}k_{x}^{\prime}}{|\omega
_{\varepsilon,+}|}}\\
\sqrt{\eta^{-1}-\frac{s_{z}\gamma\upsilon_{B}k_{x}^{\prime}}{|\omega
_{\varepsilon,+}|}}%
\end{array}
\right)  ,
\end{align}
and%
\begin{align}
\omega_{\varepsilon,-} &  =-\eta^{-3}\sqrt{2n\omega_{c}^{2}+\eta^{2}\gamma
^{2}\upsilon_{B}^{2}k_{x}^{\prime2}},\\
|\Phi_{n,-}\rangle &  =\frac{1}{\sqrt{2}}\left(
\begin{array}
[c]{c}%
\sqrt{\eta^{-1}-\frac{s_{z}\gamma\upsilon_{B}k_{x}^{\prime}}{|\omega
_{\varepsilon,-}|}}\\
-\sqrt{\eta^{-1}+\frac{s_{z}\gamma\upsilon_{B}k_{x}^{\prime}}{|\omega
_{\varepsilon,-}|}}%
\end{array}
\right)  .
\end{align}
For $n=0$, Eqs. (\ref{eq_C1})-(\ref{eq_C3}) reduce to%
\begin{equation}
|\psi_{0}\rangle=-\frac{\eta\omega_{+}}{\sqrt{2}\omega_{c}}D_{0}|0\rangle,
\end{equation}%
\begin{equation}
\hat{\Pi}|\varphi_{0}\rangle=-\frac{\lambda_{+}\lambda_{-}}{\omega_{-}}%
D_{0}\left(  2\eta^{2}\omega_{\varepsilon}\lambda_{+}|0\rangle-\omega
_{+}|1\rangle\right)  ,
\end{equation}%
\begin{equation}
\left(  \omega_{-}\hat{\Pi}+\omega_{+}\hat{\Pi}^{\dag}\right)  |\psi
_{0}\rangle=\frac{\eta^{2}\omega_{+}}{\sqrt{2}\omega_{c}}D_{0}\left(
4\eta\omega_{\varepsilon}\lambda_{+}\lambda_{-}|0\rangle-\omega_{+}%
|1\rangle\right)  .
\end{equation}
Then, we can derive for%
\begin{align}
\omega_{\varepsilon,0} &  =-\eta^{-2}s_{z}\gamma\upsilon_{B}k_{x}^{\prime},\\
|\Phi_{0}\rangle &  =\left(
\begin{array}
[c]{c}%
0\\
1
\end{array}
\right)  .
\end{align}
Finally, by extending $n$ to integers, the coefficients of the wavefunction
and the LLs can be uniformly expressed as Eqs. (\ref{eq_Rn}) and (\ref{eq_mn})
of the main texts.

\end{document}